%% file: main.tex
\documentclass[sigconf]{acmart}
\usepackage{xspace}

\AtBeginDocument{%
  \providecommand\BibTeX{{%
    \normalfont B\kern-0.5em{\scshape i\kern-0.25em b}\kern-0.8em\TeX}}}

\copyrightyear{2023}
\acmYear{2023}
\setcopyright{acmlicensed}\acmConference[COMPASS '23]{ACM SIGCAS/SIGCHI
Conference on Computing and Sustainable Societies}{August 16--19,
2023}{Cape Town, South Africa}
\acmBooktitle{ACM SIGCAS/SIGCHI Conference on Computing and Sustainable
Societies (COMPASS '23), August 16--19, 2023, Cape Town, South Africa}
\acmPrice{15.00}
\acmDOI{10.1145/3588001.3609364}
\acmISBN{979-8-4007-0149-8/23/08}

\begin{document}

\title{\systemname: Food Waste Reduction and Behavior Change on Campus with Data Visualization and Gamification}

\author{Yue Yu}
\email{yyubv@connect.ust.hk}
\orcid{0000-0002-9302-0793}
\affiliation{
  \institution{The Hong Kong University of Science and Technology}
  \state{Hong Kong SAR}
  \country{China}
}

\author{Sophia Yi}
\email{syi@connect.ust.hk}
\orcid{0009-0004-4347-0878}
\affiliation{
  \institution{The Hong Kong University of Science and Technology}
  \state{Hong Kong SAR}
  \country{China}
}

\author{Xi Nan}
\email{xnan@connect.ust.hk}
\orcid{0009-0005-0901-1296}
\affiliation{
  \institution{The Hong Kong University of Science and Technology}
  \state{Hong Kong SAR}
  \country{China}
}

\author{Leo Yu-Ho Lo}
\email{leoyuho.lo@connect.ust.hk}
\orcid{0000-0002-3660-3765}
\affiliation{
  \institution{The Hong Kong University of Science and Technology}
  \state{Hong Kong SAR}
  \country{China}
}

\author{Kento Shigyo}
\email{kshigyo@connect.ust.hk}
\orcid{0000-0002-5095-7500}
\affiliation{
  \institution{The Hong Kong University of Science and Technology}
  \state{Hong Kong SAR}
  \country{China}
}

\author{Liwenhan Xie}
\email{liwenhan.xie@connect.ust.hk}
\orcid{0000-0002-2601-6313}
\affiliation{
  \institution{The Hong Kong University of Science and Technology}
  \state{Hong Kong SAR}
  \country{China}
}

\author{Jeffry Wicaksana}
\email{jwicaksana@connect.ust.hk}
\orcid{0000-0001-8275-0996}
\affiliation{
  \institution{The Hong Kong University of Science and Technology}
  \state{Hong Kong SAR}
  \country{China}
}

\author{Kwang-Ting Cheng}
\email{timcheng@ust.hk}
\orcid{0000-0002-3885-4912}
\affiliation{
  \institution{The Hong Kong University of Science and Technology}
  \state{Hong Kong SAR}
  \country{China}
}

\author{Huamin Qu}
\orcid{0000-0002-3344-9694}
\email{huamin@cse.ust.hk}
\affiliation{
  \institution{The Hong Kong University of Science and Technology}
  \state{Hong Kong SAR}
  \country{China}
}

\renewcommand{\shortauthors}{Yu et al.}
\newcommand{\systemname}{\textit{FoodWise}\xspace}
\newcommand{\universityname}{The Hong Kong University of Science and Technology (HKUST)\xspace}
\newcommand{\shortuniversityname}{HKUST\xspace}

\begin{abstract}
Food waste presents a substantial challenge with significant environmental and economic ramifications, and its severity on campus environments is of particular concern. In response to this, we introduce \systemname, a dual-component system tailored to inspire and incentivize campus communities to reduce food waste. The system consists of a data storytelling dashboard that graphically displays food waste information from university canteens, coupled with a mobile web application that encourages users to log their food waste reduction actions and rewards active participants for their efforts.

Deployed during a two-week food-saving campaign at \universityname in March 2023, \systemname engaged over 200 participants from the university community, resulting in the logging of over 800 daily food-saving actions. Feedback collected post-campaign underscores the system's efficacy in elevating user consciousness about food waste and prompting behavioral shifts towards a more sustainable campus. This paper also provides insights for enhancing our system, contributing to a broader discourse on sustainable campus initiatives.

\end{abstract}





\maketitle

\input{sections/introduction}

\input{sections/objectives}

\input{sections/methodology}
\input{sections/evaluation}
\input{sections/discussion}
\input{sections/conclusion}

\begin{acks}
This research was supported by the project under the
Sustainable Smart Campus as a Living Lab (SSC) Initiative
from The Hong Kong University of Science and Technology
(Grant \#: F0817). The authors would like to thank Ka Wa Chiu, Yiu Tung Lam, and Ka Yiu Wong for their contribution to the development of the video processing pipeline, and Prof. Yuhan Luo and Prof. Mingming Fan for valuable feedback.
\end{acks}

\bibliographystyle{ACM-Reference-Format}
\bibliography{sample-base}

\appendix

\end{document}

%% file: sections/introduction.tex
\section{Introduction}
\label{sec:intro}

Food waste constitutes a significant environmental issue, contributing substantially to environmental degradation and hindering the efficient use of resources \cite{Stenmarck2016}. This issue is particularly pronounced within the context of college campuses. For instance, \universityname has reported an average monthly accumulation of over 40 tons of food waste generated on campus.

\begin{figure*} [t]
 \centering 
 \vspace{-0.3cm}
 \includegraphics[width=\linewidth]{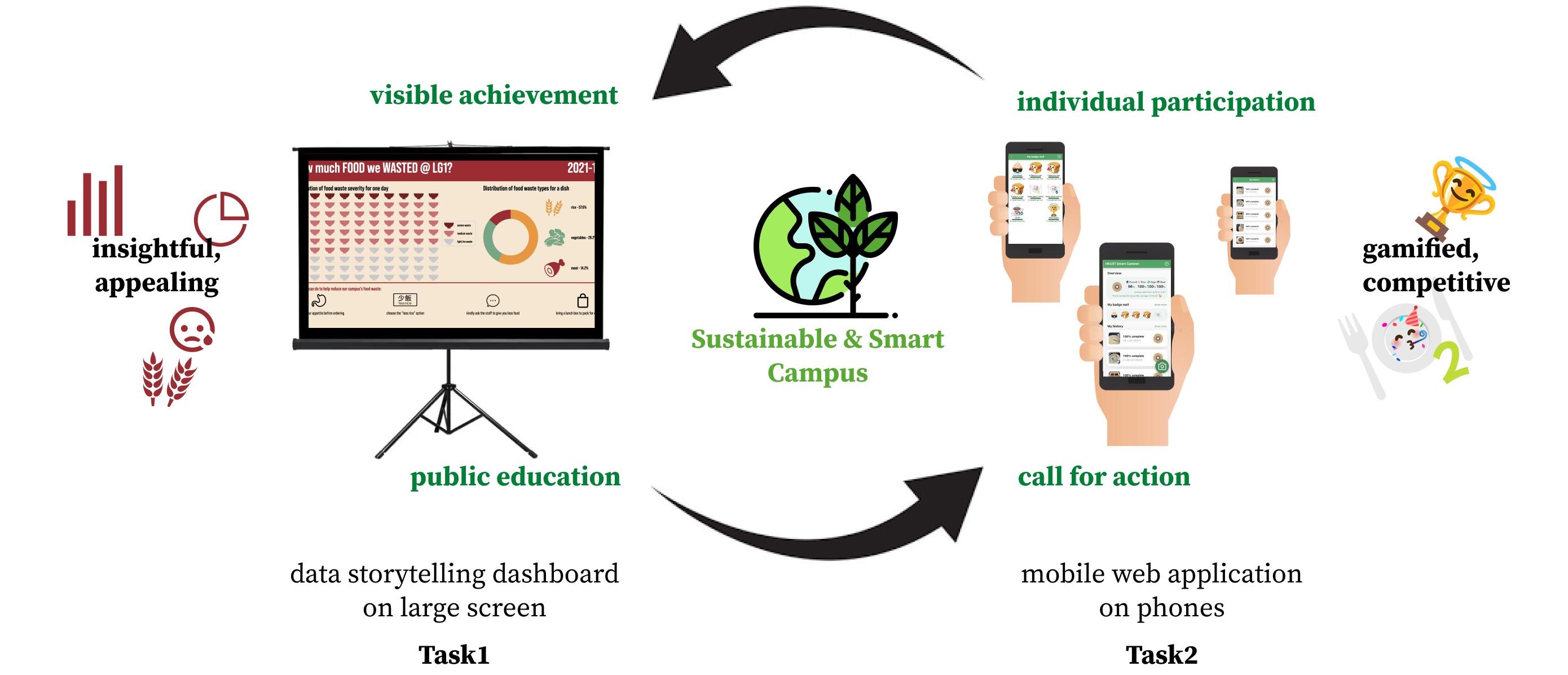} 
 \vspace{-0.3cm}
 \caption{Our proposed system consists of a data storytelling dashboard that raises awareness about the severity of campus food waste (Task 1) and a mobile web application that encourages action to reduce food waste (Task 2). Together, they form a feedback loop promoting food-saving behavior for a more sustainable campus.}
 \Description[FoodWise system overview.]{The diagram includes a large screen dashboard for public education about food waste and mobile web applications promoting waste reduction. They exchange feedback centered around a green Earth symbol, highlighting the sustainable goal.}
 \vspace{-0.2cm}
 \label{fig:overview}
\end{figure*}
Various interventions have been developed and implemented to address this issue, ranging from information-based strategies to technological solutions and policy changes \cite{reynolds2019consumption}. For example, students at Baldwin Wallace University developed ``Campus Plate'', a platform that facilitates the collection and distribution of surplus food from dining services and campus events \cite{campusplate}. Leanpath proposed a system that utilizes a weight sensor and a camera to track and analyze food waste data, offering comprehensive insights via a cloud-based dashboard \cite{leanpath}. Alipay Ant Forest introduced a gamified mobile phone application, incentivizing users to finish their meals by rewarding the ``virtual green energy'' that can be used to adopt a real tree in the desert \cite{alipay}. Additionally, social media has been leveraged as an effective tool for spreading knowledge about food waste, with studies showing its impact on raising awareness of waste generation \cite{young2017can}.

Inspired by these successful endeavors, we present \systemname, a system that leverages advanced technologies and human-centered computing to motivate sustainable behavior change. Our approach lies in the integration of a data storytelling dashboard and a mobile web application, together arousing awareness amongst the campus community about food waste issues and cultivating sustainable food-saving habits. The system was deployed in a two-week food-saving campaign at \shortuniversityname which attracted over 200 participants including students, faculties, and staffs. A post-study survey with 53 respondents validated the effectiveness of the system in heightening awareness of food waste issues and fostering food-saving behavior on campus. This system provides a blueprint for other institutions aiming to reduce food waste and foster sustainability within their communities through the strategic use of data visualization and human-centered design.

%% file: sections/objectives.tex
\section{Objectives}
\label{sec: backg}

Our study is motivated by a comprehensive review on food waste reduction interventions, which categorizes applied interventions into three types: information-based, technological solutions, and policy/system/practice change \cite{reynolds2019consumption}. Information-based interventions are designed to raise awareness and stimulate reflection on the food waste problem. Technological solutions employ technologies to track food waste data and promote food-saving behavior. Policy/system/practice change interventions advocate for larger-scale changes to the way food is managed and consumed.

While policy/system/practice change interventions often require substantial institutional support and changes to existing infrastructure, which may not be feasible within the scope of our current project, we strategically chose to focus on the first two intervention types, information-based and technological solutions. These interventions align well with our campus environment, where we can leverage existing technological infrastructure and the inherent openness of the university community to new information and behavioral changes.
Therefore, the primary objective of our system is to foster food-saving habits of the university community members and contribute to building a sustainable campus by leveraging these two intervention types. To achieve this goal, we have divided it into two actionable tasks:

\noindent\textbf{Task 1: Raising awareness about the severity of campus food waste}:
In alignment with the information-based intervention strategy, we propose a large-screen data storytelling dashboard that vividly depicts the food waste information on campus. The dashboard is designed to provide insights to the audience and stimulate reflection on the existing food waste problem. Consequently, the interface needs to be visually appealing yet capable of eliciting negative emotions to stir environmental consciousness about food wastage.

\noindent\textbf{Task 2: Encouraging action to reduce food waste at every meal}:
In line with the technological solutions intervention strategy, we propose a mobile web application that motivates users to record their food waste reduction efforts and track the changes in their behavior throughout a food-saving campaign, fostering long-term food conservation habits. To enhance engagement, we incorporate gamification elements and reward active participants.

As depicted in Figure \ref{fig:overview}, our proposed system comprises two main components: a data storytelling dashboard on a large screen and a mobile web application. The dashboard presents data-driven insights to raise food waste awareness among the HKUST community. Concurrently, the gamified web application and the food-saving campaign motivate participants to reduce food wastage in exchange for rewards. The individual efforts are then collectively displayed on the dashboard, creating a feedback loop that encourages continued food waste reduction. This approach aligns with our ultimate objective of cultivating a sustainable and smart campus environment.

%% file: sections/methodology.tex
\begin{figure} [t]
 \centering 
 \vspace{0cm}
 \includegraphics[width=\linewidth]{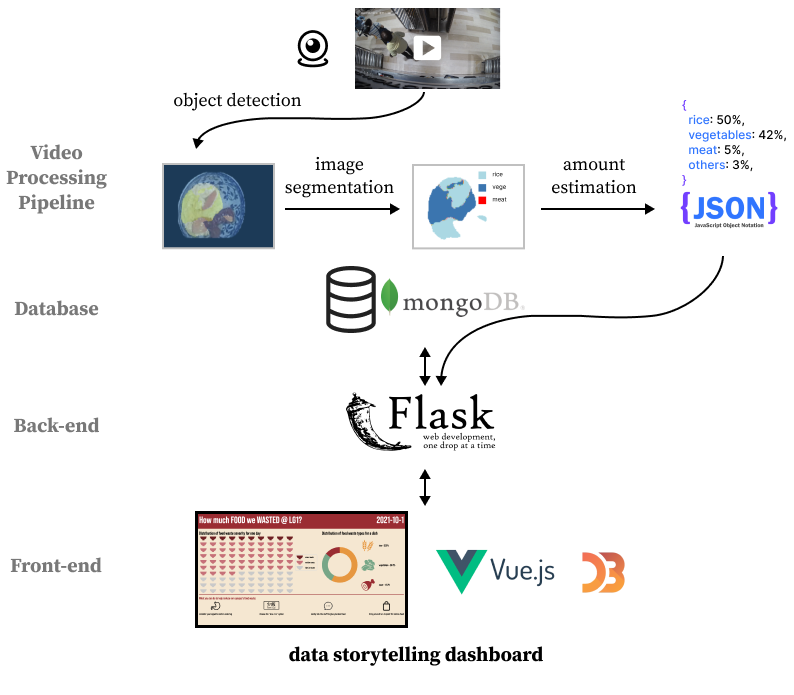} 
 \vspace{-0.3cm}
 \caption{
     The system architecture of the dashboard featuring four key components: a video processing pipeline that quantifies different types of food waste from video clips, a structured data storage database, a back-end system managing data requests, and a front-end data visualization website.
 }
 \vspace{0cm}
 \Description{System architecture diagram illustrating four layers of the dashboard. The top layer is the video processing pipeline, where footage from the canteen camera is processed and quantified into various types of food waste. The processed data is stored in the second layer, the database. The third layer, the back-end system, manages data requests and interacts with the fourth layer, the front-end, which visualizes the data on the website. Each layer interacts systematically with adjacent layers to provide an efficient flow of information.}
 \label{fig:dashboard_arch}
\end{figure}
\section{Method}
\label{sec: method}

In this section, we provide a thorough overview of \systemname, which consists of a data storytelling dashboard designed to raise public consciousness and a mobile web application built to promote user participation. Following this, we present a campus-wide food-saving campaign that incorporates \systemname. Collectively, these elements form a multifaceted intervention strategy, aiming to foster sustainable food consumption behaviors.

\subsection{Data Storytelling Dashboard}

The architecture of the dashboard consists of four main modules, as shown in Figure \ref{fig:dashboard_arch}:
\begin{itemize}
    \item A video processing pipeline that processes food waste information captured from video clips.
    \item A database to store structured data.
    \item A back-end system that handles data requests from the front-end via application programming interfaces (APIs).
    \item A front-end data storytelling website designed for data visualization and analytics, targeting the public audience.
\end{itemize}

The video processing pipeline analyzes the surveillance footage captured by the tray return station cameras in a canteen on campus. It aims to estimate the quantities of three types of food waste, including rice, meat, and vegetables. These categories were chosen in accordance with the food culture of the region where \shortuniversityname is located, Hong Kong, where rice is a staple food and meat and vegetables are common components of meals. This pipeline leverages several pre-trained deep-learning models.
Firstly, an object detection model trained with the YOLO architecture \cite{redmon2016yolo} is utilized to extract tray images from the video clips. Subsequently, an image segmentation model trained with the DeepLabv3 architecture \cite{chen2018deeplab} is employed to segment and identify the various types of food waste present. Finally, the areas occupied by each type of food waste on each tray are stored in JSON format within a MongoDB database \cite{mongo}.

For the back-end, we utilized Flask \cite{flask}, a Python-based framework. The back-end interacts with the front-end to retrieve daily and monthly aggregated statistics. To ensure efficiency, we designed a scheduler that runs on a daily basis. This scheduler triggers various statistical algorithms and caches the most recent aggregate data in the database for quick retrieval. Additionally, we developed a regression model for estimating the weight of food waste based on the pixel area. The dependent variable in this analysis is the weight data obtained from the university database, while the calculated pixel area serves as the independent variable.

\begin{figure} [t]
 \centering 
 \vspace{0cm}
 \includegraphics[width=\linewidth]{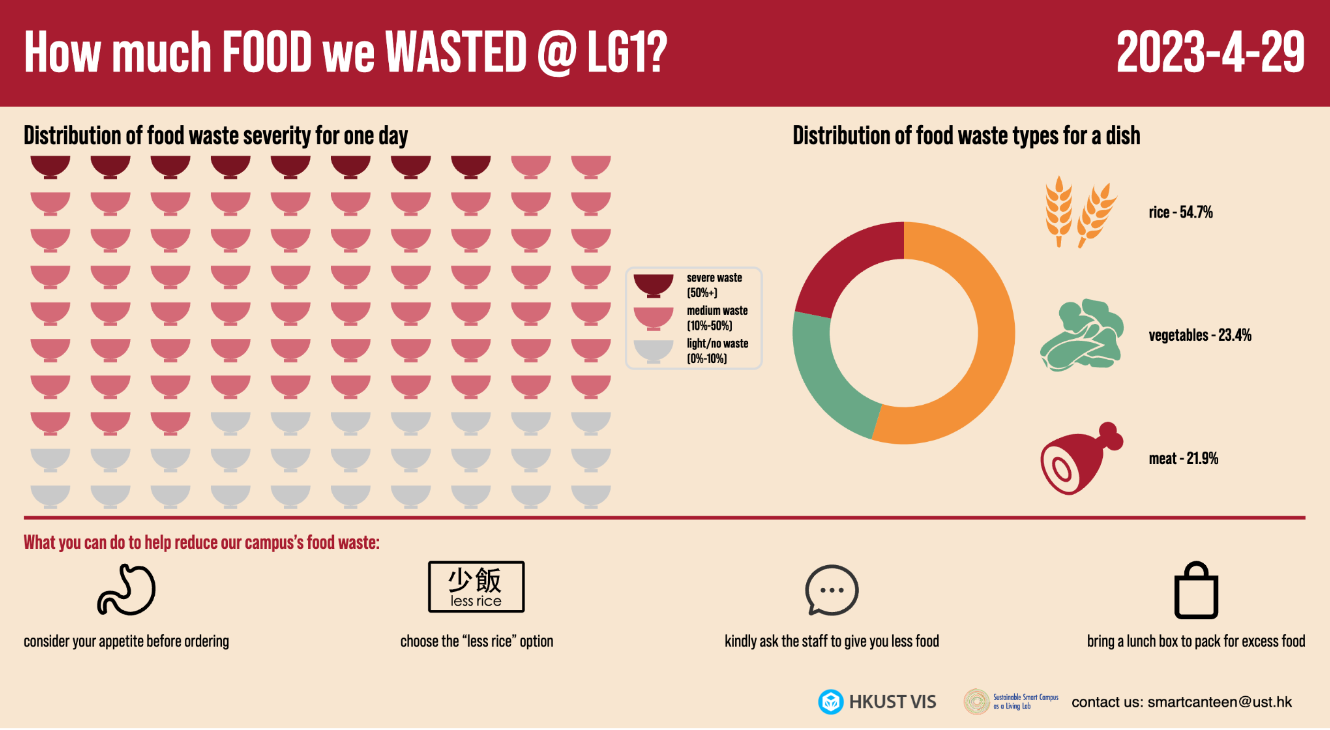} 
 \vspace{-0.3cm}
 \caption{
     The Daily page of the dashboard, featuring data visualization and graphic illustrations to represent the severity and types of food waste, updated in real-time.
 }
 \Description{The "Daily" dashboard page features two main charts. A header states "How much food we wasted?" with the current date underneath. In the body, the left shows a distribution chart using 100 bowls to visualize the severity of food waste, with three severity levels indicated by varying intensities of red color: severe, medium, and light. The right side displays a ring chart dividing food waste into three types: rice, meat, and vegetables. The footer presents four tips with accompanying illustrations as reminders to conserve food.}
 \vspace{0cm}
 \label{fig:dashboard_ui1}
\end{figure}

\begin{figure} [!htb]
 \centering 
 \vspace{0cm}
 \includegraphics[width=\linewidth]{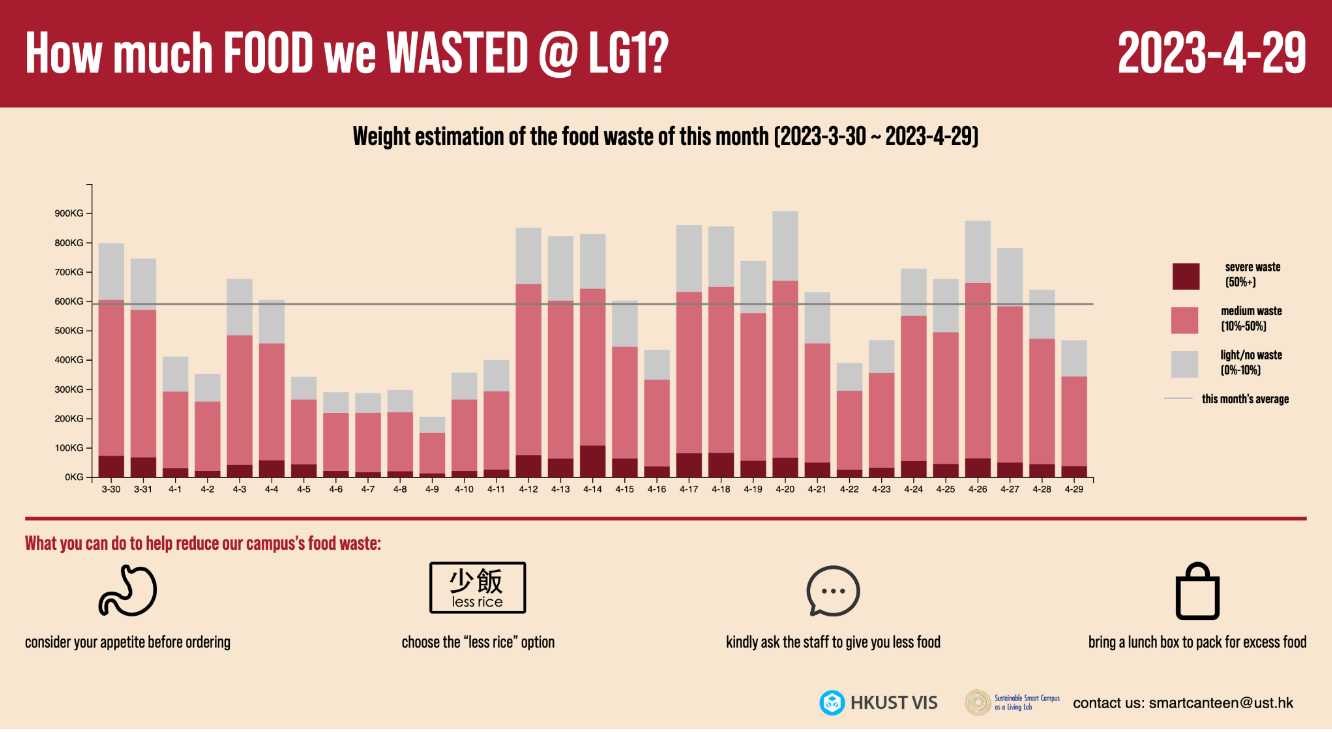} 
 \vspace{-0.3cm}
 \caption{
     The monthly page of the dashboard, showcasing bar charts to illustrate the changing trend of food waste on campus over time, updated monthly.
 }
 \Description{The graph displays a monthly snapshot of food waste on campus in a stacked bar chart format. The header shows the question 'How much food we wasted?' followed by the date. In the body, each day of the month is represented by a vertical bar composed of three segments: severe waste, medium waste, and light waste. These segments are differentiated by varying shades of colors, with the overall height of each bar reflecting the total amount of food waste for the respective day.}
 \vspace{0cm}
 \label{fig:dashboard_ui2}
\end{figure}

To enhance the emotional impact of the front-end, we referred to academic research on emotion elicitation in data stories \cite{Lan21smile, lan2022negative} and incorporated some relevant design heuristics into our dashboard design. Using Vue.js \cite{vue}, we developed two pages that aim to evoke an emotional response from the audience.
The first page, updated daily, displays the severity and type of food waste using data visualization and graphic illustrations (Figure \ref{fig:dashboard_ui1}). 
We visually depict the extent of waste distribution using 100 bowls, a symbolism deeply rooted in the local food culture, where each bowl stands for 1\% of the total meals.
By presenting timely and relatable information, this page seeks to resonate with the audience and capture their attention.
The second page, updated monthly, showcases a series of bar charts illustrating the changing trend of food waste on campus (Figure \ref{fig:dashboard_ui2}). This provides a broader perspective on the issue and allows users to track progress over time.
To make the dashboard visually appealing on large screens, we incorporated animations to enhance the presentation of the data visualization components with the two pages seamlessly transitioning in an infinite loop.
In addition to providing data stories, we aimed to prompt action from the audience. Therefore, each page includes food-saving tips such as ``consider your appetite before ordering'', ``choose the `less rice' option'', ``kindly ask the staff to give you less food'', and ``bring a lunch box to pack excess food''. By incorporating these tips, we encourage users to take practical steps toward reducing food waste.

\subsection{Mobile Web Application}
\begin{figure} [!htb]
 \centering 
 \vspace{0cm}
 \includegraphics[width=\linewidth]{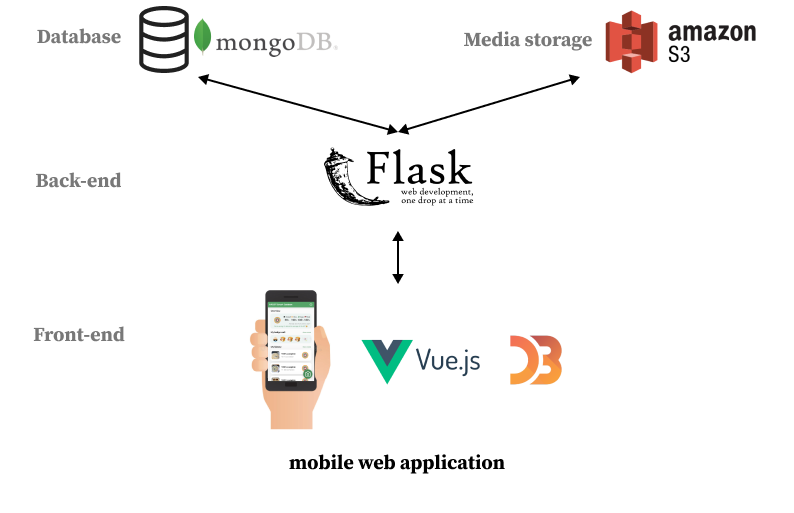} 
 \vspace{-0.3cm}
 \caption{
     System architecture of the mobile web application, comprising a user database, file storage service, back-end APIs, and a front-end mobile web application.
 }
 \Description{Mobile web application architecture layered as database and media storage, back-end, and front-end. The back-end layer acts as a bridge, storing and retrieving data from the database and media storage to be displayed on the front-end.}
 \vspace{0cm}
 \label{fig:app_arch}
\end{figure}

We developed a mobile web application to encourage self-tracking and reflective behaviors~\cite{luo2021foodscrape}, where we employed gamified design to engage participants.
The web application's architecture, shown in Figure \ref{fig:app_arch}, consists of four main modules:
\begin{itemize}
    \item A database for user data storage.
    \item A storage service for user-uploaded files.
    \item A back-end system offering functions via application programming interfaces (APIs).
    \item A front-end web application on mobile phones providing various user features.
\end{itemize}

Like the dashboard, the web application's database is MongoDB-based. We utilize AWS Simple Storage Service (S3) \cite{s3} for file storage. The back-end interacts with the file storage system and database to facilitate various functions. It allows users to register, log in, upload photos and food completion percentages, view historical data, and engage in gamification activities.

\begin{figure} [!htb]
 \centering 
 \vspace{0cm}
 \includegraphics[width=\linewidth]{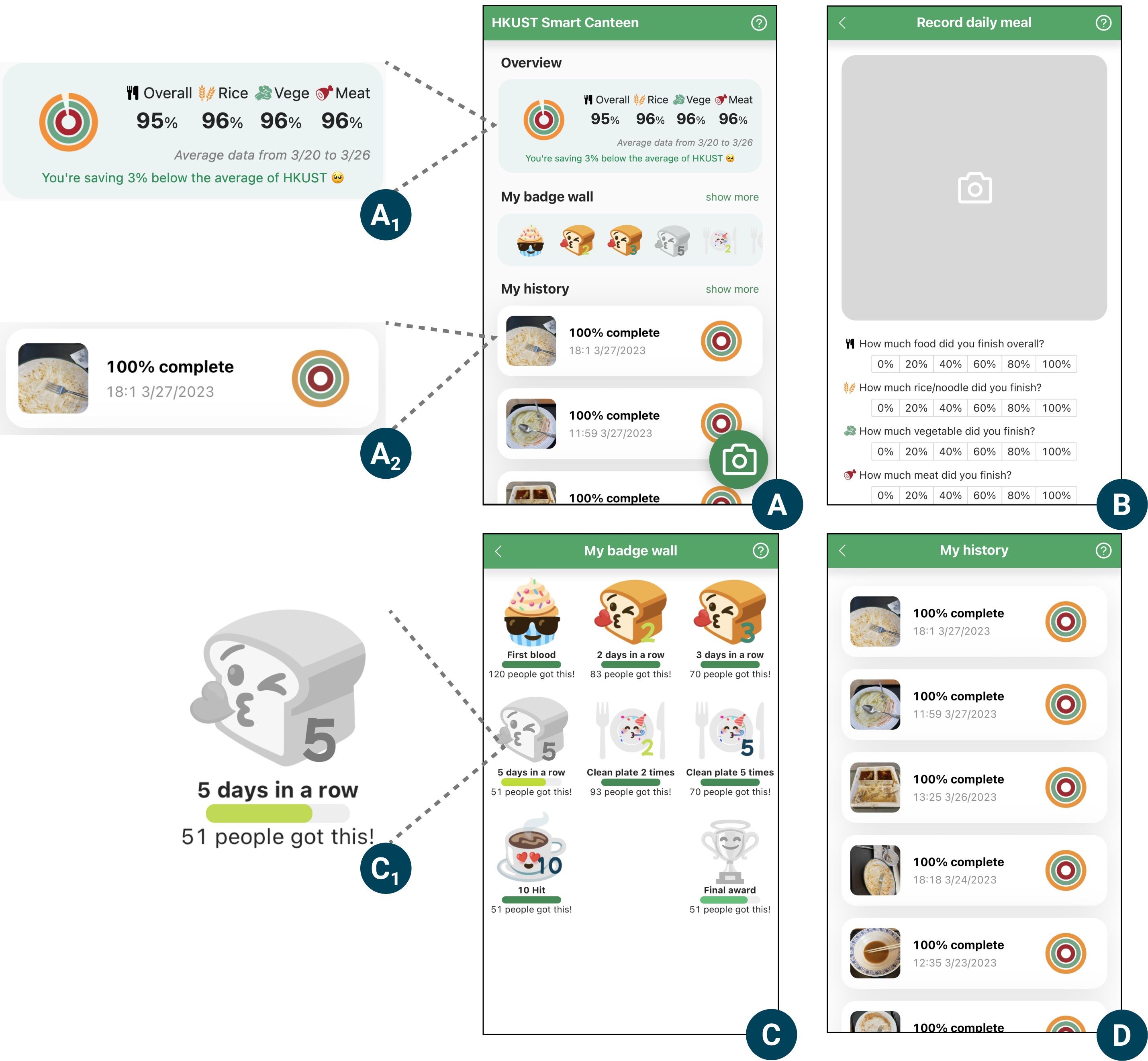} 
 \vspace{-0.3cm}
 \caption{
     User interface of the mobile web application: (A) the Overview page with user statistics and comparison with other users, (B) the Record page for logging food-saving actions, (C) the Badge page showing progress towards earning badges, and (D) the History page for reviewing past food-saving records.
 }
 \Description{
     A snapshot of a mobile web application consisting of four key sections. Section A is the Overview page, portraying user's food-saving statistics compared to others. It has a data panel on top, exhibiting the user's average completion statistics and the community average. A ring chart encodes the completion percentage, and badges earned are shown in the middle, with past submissions and their completion ring chart at the bottom. A camera button at the bottom right allows for quick navigation to the Record page. Section B is the Record page, where users can photograph their finished meals and input their food completion scores. Section C, the Badge page, visualizes the earned badges, accompanied by a progress bar showing the remaining effort required. It also displays the count of users who have earned each badge. Section D is the History page, giving users a list of food-saving records.
 }
 \vspace{0cm}
 \label{fig:app_ui}
\end{figure}

For the front-end, we prioritized a web-based application over native iOS or Android applications due to its advantages in cross-platform compatibility, maintenance efficiency, and accelerated development cycle. We utilized Vue.js to build a versatile web application, as displayed in Figure \ref{fig:app_ui}.
The Overview page (Figure \ref{fig:app_ui}-A) presents a comprehensive summary of users' self-recorded food-saving actions. In this context, food-saving actions refer to the process of users uploading photos of their finished meals and self-reporting the completeness score for three types of ingredients. These ingredient types align with the food waste categories displayed on the storytelling dashboard, namely rice, vegetables, and meat. Initially, we also explored the integration of computer vision technology to automatically quantify leftovers on a plate, drawing inspiration from the video processing algorithm employed in the data storytelling dashboard. However, our pilot study revealed that this approach could lead to laggy responses and potential inaccuracies, compromising user experience. As a result, we eventually decided to rely on the honesty of user-reported values from our community members. The data panel (Figure \ref{fig:app_ui}-A1) at the top displays individual user's average completion statistics across different types of food. A ring chart on the left visually encodes the completion percentage, using arcs of varying lengths to represent different completion rates. This panel also juxtaposes user statistics with average completion values of all registered users, enabling users to compare their performance against the broader campus community. The middle section of the page exhibits the array of badges earned by users. At the bottom, users can review their past submissions along with the corresponding completion ring chart (Figure \ref{fig:app_ui}-A2). A conveniently located camera button at the bottom right allows users to navigate swiftly to the Record page (Figure \ref{fig:app_ui}-B) and log their food-saving actions post-meal.
The Record page (Figure \ref{fig:app_ui}-B) guides users to take a photo of their finished meals and input their completion score of each type of food. Each submission of this record is considered a food-saving action and contributes to the user's progress towards earning badges and final rewards.
The Badge page (Figure \ref{fig:app_ui}-C) displays badges earned, supplemented by a progress bar (Figure \ref{fig:app_ui}-C1) that indicates the remaining effort required to attain each badge. This page also showcases the number of users who have earned each badge, fostering a competitive spirit and motivating users to strive for these badges.
Lastly, the History page (Figure \ref{fig:app_ui}-D) provides users with access to their record list, offering a retrospective view of their food-saving journey.

\begin{figure} [t]
 \centering 
 \vspace{0cm}
 \includegraphics[width=\linewidth]{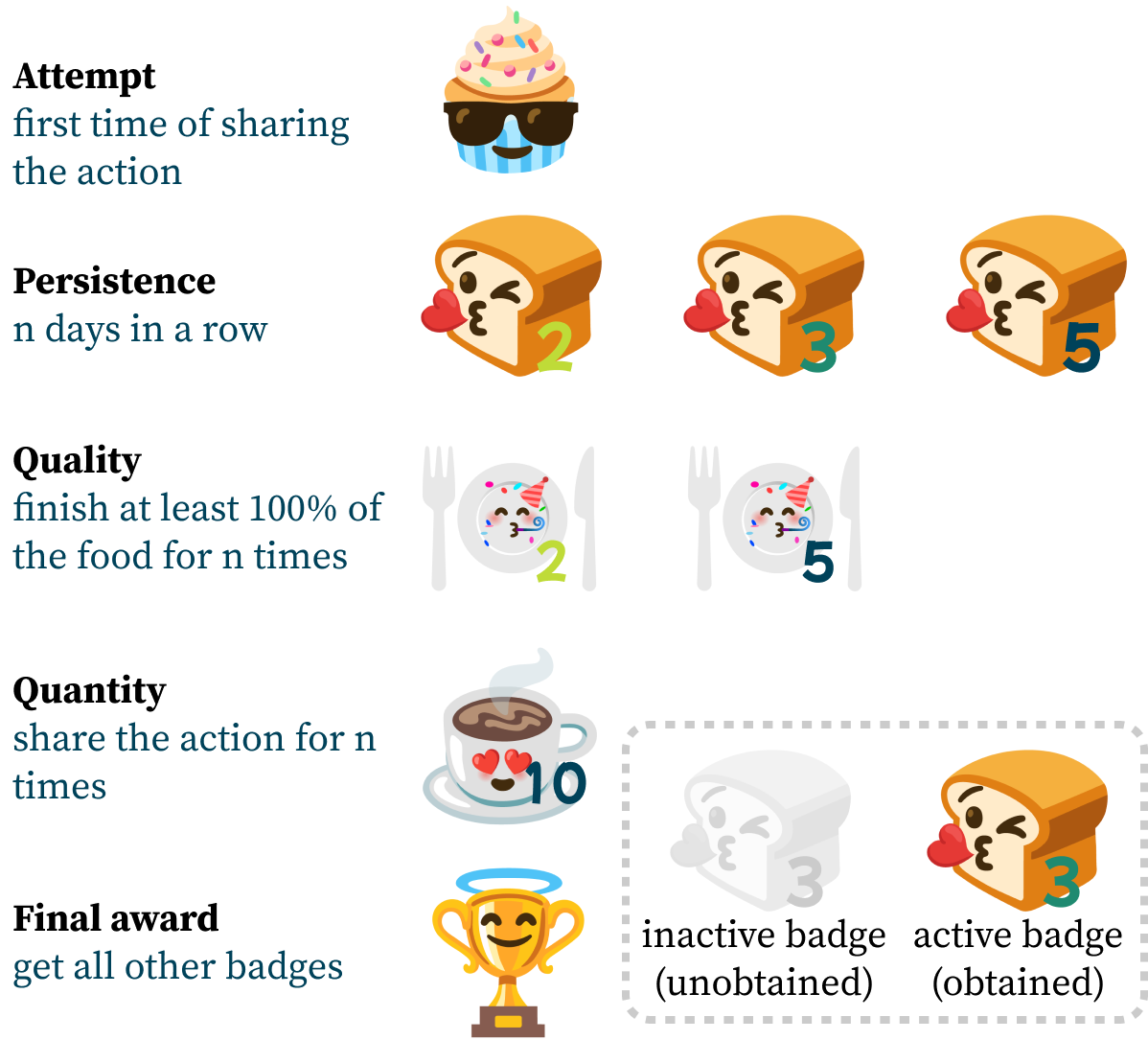} 
 \vspace{-0.3cm}
 \caption{
     Gamification elements integrated into the web application, including badges awarded for different food-saving goals to enhance user engagement.
 }
 \Description{
    Cartoon food badges encouraging food conservation. The diagram presents four types of badges symbolizing different levels of engagement: attempt, persistence, quantity, and quality. Each badge is represented by a distinctive cartoon food icon - such as bread, cake, or coffee. Badges are colorless when inactive while colorful when active.
 }
 \vspace{0cm}
 \label{fig:badges}
\end{figure}

The web application integrates a gamification feature to maintain user engagement through competition and goal-setting. Gamification has proven to be effective in increasing user engagement and promoting behavior change \cite{deterding2011game}. As shown in Figure \ref{fig:badges}, this long-term, habit-forming challenge is made more interesting by setting different levels of goals in the campaign, including attempt, persistence, quantity, and quality. Each time users achieve a goal, they earn a badge accordingly. The multi-tiered badge system creates a playful atmosphere by encouraging participants to make meaningful choices on their own \cite{nicholson2015recipe}.
The Attempt badge is awarded upon the user's first successful submission. The Persistence badge requires users to record actions for several consecutive days, encouraging a daily food-saving habit. Quality and Quantity badges motivate users to consistently finish meals instead of making several random attempts for rewards. To claim the final reward, users must earn all of the badges.

\subsection{Campaign Organization}
To generate positive social impact and evaluate the real-world effectiveness of \systemname, we incorporated our system into a two-week food-saving campaign named ``Save Food, Win Free Meals'' in March and April 2023. Participants who acquired all the badges were rewarded with a coupon that is worth approximately \$6.5 from a local healthy food store, sufficient for one free meal. The campaign timeline commenced with a pre-registration period from March 13 to March 19 to engage potential participants. The campaign officially took flight from March 20 to April 3, spanning over two weeks.

\begin{figure} [t]
 \centering 
 \vspace{0cm}
 \includegraphics[width=\linewidth]{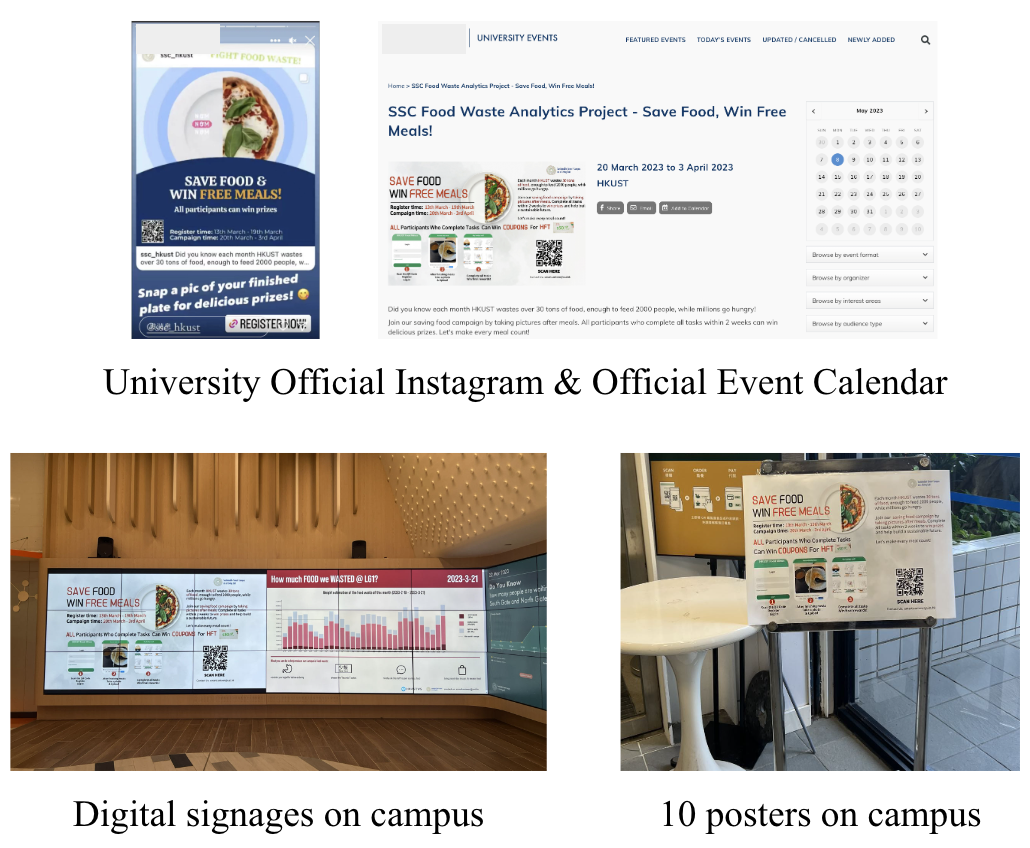} 
 \vspace{-0.3cm}
 \caption{
     The promotion channels of the food-saving campaign on campus.
 }
 \Description{
    The promotion channels of the food-saving campaign on campus. The figure showcases four main methods of communication: the university's Instagram post and digital calendar featured at the top, a dashboard with the event's poster on the campus LED screens, and a printed poster displayed on signage captured at the lower part of the image.
 }
 \vspace{0cm}
 \label{fig:promotion}
\end{figure}

To promote participation, we developed promotional materials, comprising printed posters and digital content for social media platforms. With the aid of the Sustainable Office at \shortuniversityname, we employed various channels to publicize the campaign, as illustrated in Figure \ref{fig:promotion}. This included sharing campaign details on \shortuniversityname's official Instagram with about 20k followers, posting the event on the university event calendar, featuring the poster and real-time data storytelling dashboard on digital signages on campus, and displaying posters at 10 high-visibility locations on campus, particularly around the canteens.

Our campaign successfully attracted 220 users who registered on the web application. Throughout the campaign, users logged a total of 811 food-saving actions, and 51 users won the final award.

%% file: sections/evaluation.tex
\begin{figure*} [!htb]
 \centering 
 \vspace{0cm}
 \includegraphics[width=\linewidth]{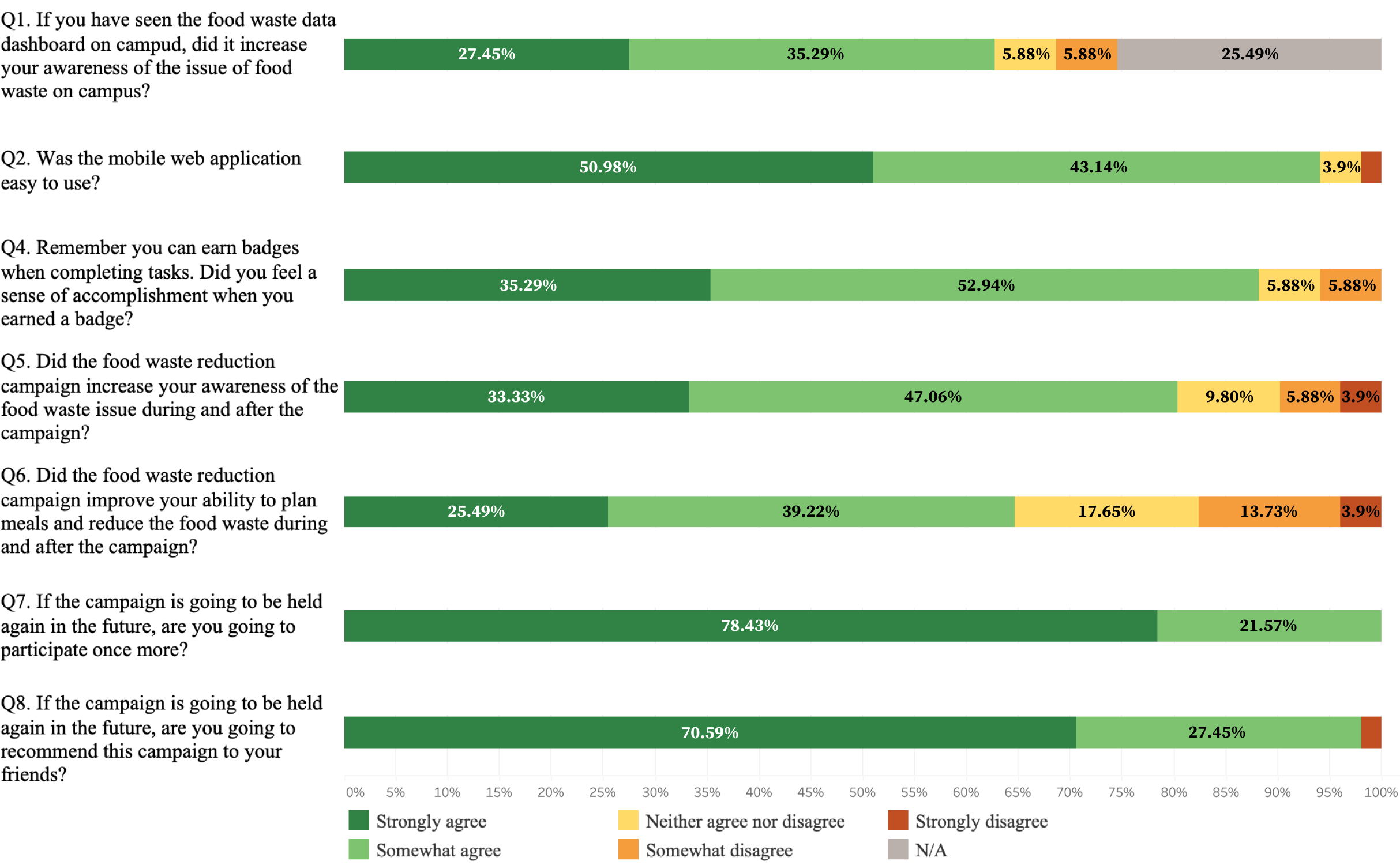} 
 \vspace{-0.3cm}
 \caption{
     Post-study user survey responses (N=53) on a 5-point Likert scale evaluating the large-screen data storytelling dashboard's role in raising food waste awareness (Q1), the usability and impact of the mobile web application (Q2, Q4), the influence of the campaign on changing user behavior and promoting food-saving habits (Q5, Q6), and the potential interest in future food-saving campaigns (Q7, Q8).
 }
    \Description{
    The responses to the survey are illustrated on a 5-point Likert scale that ranges from "strongly disagree" to "strongly agree". The results vary across questions, but a pattern of positive responses can be noticed across all areas. }
 \vspace{0cm}
 \label{fig:eva_likert}
\end{figure*}

\section{Evaluation}
\label{sec:evaluation}

We conducted a post-campaign survey for a comprehensive evaluation of the effectiveness of our system and the campaign. To encourage user feedback, we offered extra five coupons worth 50 Hong Kong dollars ($\sim$\$6.5) through a lucky draw for survey respondents. We received responses from 53 participants. The questionnaire consists of 7 single-answer Likert scale questions (Figure \ref{fig:eva_likert}), 1 multiple-answer question (Figure \ref{fig:eva_issue}), and 1 open-ended question. Our usability-related questions were informed by the Post-Study System Usability Questionnaire (PSSUQ), a widely used tool for evaluating the usability of systems \cite{lewis2002psychometric}.

\subsection{Dashboard Evaluation (Task 1)}\label{dashboard_eva}
Question 1 (Q1) asked whether the large-screen data storytelling dashboard increased users' awareness of campus food waste. Over 60\% of respondents agreed with its role in heightening their awareness. However, about a quarter of respondents reported ``N/A'', meaning they have not noticed the dashboard, suggesting that its placement could be optimized for better visibility.

\subsection{Web Application Evaluation (Task 2)}

\begin{figure} [t]
 \centering 
 \vspace{0cm}
 \includegraphics[width=\linewidth]{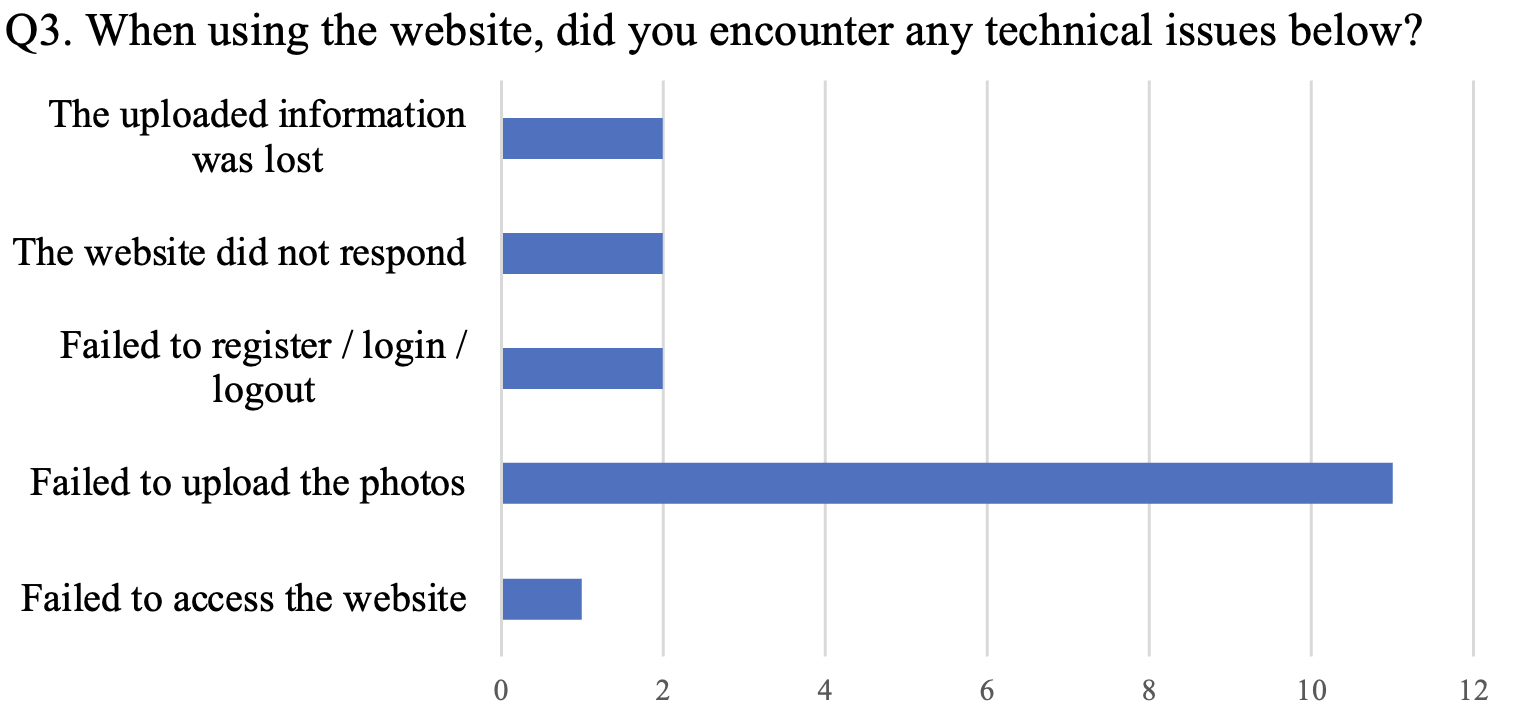} 
 \vspace{-0.3cm}
 \caption{
     Technical issues encountered in the web application (Q3), with 37 respondents reporting ``N/A''.
 }
 \Description{
    A bar chart illustrates the distribution of the answers to question 3. The "Failed to upload the photos" category has the longest bar, with eleven answers, and other categories are all with less than three answers.
 }
 \vspace{0cm}
 \label{fig:eva_issue}
\end{figure}

We evaluated the usability of our mobile web application through Q2, with over 90\% finding the application user-friendly. Figure \ref{fig:eva_issue} displays the responses to Q3, revealing that approximately 70\% of users experienced no technical issues. However, the prevalent issue among the remaining respondents was the difficulty in uploading photos, indicating a potential area for enhancement, such as implementing a front-end image compressor.

We further asked whether users felt encouraged to save food and use the application. Almost 90\% of respondents in Q4 reported a sense of accomplishment when earning a badge, suggesting it may motivate ongoing food-saving actions.

\subsection{Campaign Evaluation (Task 1 + Task 2)}

We aimed to measure the impact of the campaign on changing user behavior and promoting food-saving habits through Q5 and Q6. In Q5, more than 80\% of respondents agreed that the campaign heightened their awareness of food waste issues. Moreover, approximately 65\% of respondents in Q6 reported improved meal planning and waste reduction during and after the campaign, highlighting the campaign's effectiveness in instigating behavioral change for a more sustainable campus.

We also explored the potential of conducting similar future food-saving campaigns in Q7 and Q8. The response to Q7 underscored a strong participant interest in future campaigns, and Q8 revealed the majority's willingness to involve their friends in these initiatives. This enthusiasm suggests promising potential for expanding these campaigns, inviting a larger and more diverse participant group on campus.

\subsection{User suggestions}

To gain further insights into the user experience and campaign effectiveness, we offered an open-ended section in Q9 for participants to provide their personal thoughts. We conducted a thematic analysis of these comments, identifying common areas of improvement suggested by users, which are listed below.

\begin{itemize}
    \item \textbf{Reminder System Enhancements} Eleven users expressed forgetting to record their meal, indicating a need for a more robust reminder system. \textit{``Add notification, sometimes forget to record''} and to add \textit{``reminder notification''} were common suggestions. Incorporating these features could help users maintain a consistent routine of logging their meals.
    \item \textbf{Diversification and Flexibility of Rewards} Eight respondents advocated for a more varied range of rewards. Suggestions included coupons for cafes, cash coupons for groceries, and more flexibility in rewards overall.
    \item \textbf{Increased Campaign Awareness} Five respondents suggested improving campaign promotion and awareness, echoing the issue mentioned earlier in Section \ref{dashboard_eva}. \textit{``More eye-catching poster and promotion''} and \textit{``more information to students''} were some of the suggestions put forth. Enhancing awareness efforts could attract more participants and further elevate the campaign's impact.
    \item \textbf{Campaign Duration and Rules Clarity} Three respondents found campaign rules unclear and felt the campaign duration was too short. One user candidly shared, \textit{``The badges are too difficult for 5 days in a row because some students may not be able to eat at school every day.''} Extending the campaign and clearly communicating the rules could make it more inclusive and achievable for all participants.
\end{itemize}

%% file: sections/discussion.tex
\section{Discussion}

\subsection{Significance}

Our study showed positive outcomes, with the data storytelling dashboard and web application successfully raising food waste awareness and promoting behavior changes within a campus environment. The dashboard notably heightened awareness for 60\% of the respondents, highlighting the potential of persuasive techniques for promoting sustainability among university community members, who can act as significant change-makers due to their exposure to innovative ideas. User-friendliness and the absence of technical issues in the web application were also acknowledged with 90\% and 70\% positive feedback, respectively. The gamification feature in the web application was impactful, motivating nearly 90\% of the respondents, which suggests playful competition can be an effective strategy for promoting sustainable behavior in a community.

The broader impact of our campaign was evident with more than 80\% of respondents reporting increased food waste awareness and about 65\% confirming improved meal planning and waste reduction practices. This demonstrates the campaign's success in not only raising awareness but also instigating tangible behavior change. Furthermore, respondents' high interest in future campaigns and their willingness to promote these initiatives highlight the power of community involvement in sustainability campaigns.

\subsection{Lessons Learned}

For future organizers planning sustainability initiatives in a campus setting, we suggest several considerations to optimize their outcomes. 
Firstly, when designing the computing system, it is crucial to incorporate cultural context to create a more relevant and effective system that resonates with the local community. For example, it is recommended to carefully consider and identify the major components of food waste in the local food culture when creating an image dataset for image segmentation model training. Additionally, it is important to ensure that the visual elements used for emotion elicitation align with the local conventions and cultural norms. 
Secondly, due to the busy nature of campus life, particularly for students with rigorous schedules, an efficient yet user-friendly reminder system is crucial to maintain consistent user engagement~\cite{wu2018pulse}. For instance, the web application could request users' routine meal times and consent during registration. Consequently, personalized reminder emails aligning with meal times can be sent to enhance the likelihood of engagement.
Thirdly, while the success of the gamification elements in our campaign underscores their potential for boosting engagement, it's vital to tailor these elements to suit diverse user preferences, ensuring broader participation. For example, users making incremental progress in waste reduction could be awarded badges, even if they haven't completed all tasks. This approach could encourage sustained progress and ensure broader participation.
Lastly, while an emotionally engaging data storytelling dashboard is effective, its location in a busy campus environment is crucial. Strategic placement in high-traffic locations or even on the food ordering machine could ensure its visibility, thereby maximizing engagement among campus members. 

Additionally, while our campaign successfully engaged a portion of the campus community, the number of participants represented only about one percent of the total campus population. This highlights the importance of exploring the potential for integrating this system into broader policy changes or system-level interventions as outlined in Reynolds et al.~\cite{reynolds2019consumption}. This integration may significantly expand the system's influence and sustain long-term food-saving behavior beyond the duration of the campaign. 
For instance, future work could explore partnerships with campus dining services to integrate the system into their operations, such as incorporating food waste tracking into the meal ordering process. Additionally, collaborations with campus administration could lead to policy changes that incentivize food-saving behavior, such as discounts for students who consistently demonstrate low food waste. By integrating our system into broader institutional practices and policies, we could reach a larger audience and have a more substantial impact on food waste reduction at the campus level.

In summary, our study demonstrates an effective approach to fostering sustainable change in a campus setting by leveraging information-based and technological interventions. While we've seen promising results, continuous refinement based on user feedback and exploration of strategies for long-term engagement and policy integration are crucial. The insights from this project can inform future efforts in similar contexts, contributing to a larger goal of promoting global sustainability.

%% file: sections/conclusion.tex
\section{Conclusion}
\label{sec:concl}

We have successfully developed and deployed an innovative system named \systemname, consisting of a data-driven dashboard and a mobile web application, to encourage sustainable practices among our campus community. In a real-life scenario, our system played a central role in a highly successful two-week food waste reduction campaign, which attracted more than 200 participants. After the campaign concluded, we gathered feedback from 53 users, and the results emphasized the effectiveness of both the system and the campaign in raising awareness about food waste issues and cultivating food-saving habits among the respondents. Meanwhile, the user feedback highlighted valuable areas for improvement and offered crucial insights that will guide the enhancement of our system in future iterations.

In our ongoing pursuit of fostering sustainable habits, we recognize the significance of continuously iterating and adapting our system to meet the evolving needs of our users and address environmental challenges. Furthermore, we acknowledge the importance of integrating the system into broader policy-level interventions to sustain long-term food-saving behavior and have a more substantial impact on the campus community. We hope that our efforts will resonate beyond our campus, inspiring other communities to adopt similar initiatives and contribute to the global goal of sustainability.